# Control of charging in resonant tunneling through InAs nanocrystal quantum dots

David Katz, Oded Millo[a]

*Racah Institute of Physics, The Hebrew University, Jerusalem 91904, Israel*

Shi-Hai Kan, and Uri Banin[b]

*Department of Physical Chemistry and the Farkas Center for Light Induced Processes, The Hebrew University, Jerusalem 91904, Israel*

## ABSTRACT

Tunneling spectroscopy of InAs nanocrystals deposited on graphite was measured using scanning tunneling microscopy, in a double-barrier tunnel-junction configuration. The effect of the junction symmetry on the tunneling spectra is studied experimentally and modeled theoretically. When the tip is retracted, we observe resonant tunneling through the nanocrystal states without charging. This is in contrast to previous measurements on similar nanocrystals anchored to gold by linker molecules, where charging took place. Charging is regained upon reducing the tip-nanocrystal distance, making the junctions more symmetric. The effect of voltage distribution between the junctions on the measured spectra is also discussed.

i) E-mail: milode@vms.huji.ac.il

b) E-mail: banin@chem.ch.huji.ac.il

Tunneling spectroscopy of single semiconductor nanocrystal quantum dots (QDs) is of interest from both fundamental and applied aspects. On the one hand, the tunneling spectra provide information on the QD discrete electronic level structure,[1-4] while on the other hand, semiconductor QDs are potential building-blocks in future electronic and opto-electronic nano-devices.[5,6] The tunneling spectroscopy experiments, as well as device applications, are commonly performed in the double-barrier tunnel-junction (DBTJ) configuration, where the QD is placed between two macroscopic electrodes.[7] In this geometry, in addition to the QD level structure, the parameters of both junctions, in particular the capacitance and tunneling rate (inversely proportional to the tunneling resistance), strongly affect the I-V or dI/dV vs. V tunneling spectra characteristics. Therefore, a detailed understanding of the role played by the DBTJ geometry and the ability to control it are essential for the correct interpretation of tunneling characteristics of semiconductor QDs, as well as for their implementation in electronic nano-architectures, as demonstrated by Su et al. for semiconducting quantum wells.[8]

Scanning tunneling microscopy (STM) is an effective tool for the study of tunneling transport through metallic,[7] molecular[9] and semiconductor QDs[1-3,10] in the DBTJ configuration. Here, a DBTJ is realized by positioning the STM tip over the QD, as portrayed by the insets in Fig. 1. The capacitance and tunneling resistance of the tip-QD junction ($C_1$ and $R_1$) can be easily modified by changing the tip-QD distance, usually through the control over the STM bias and current settings ($V_s$ and $I_s$). On the other hand, the QD-substrate junction parameters ($C_2$ and $R_2$) are practically stable for a specific QD. By varying $C_1$ one can modify the single electron charging energy, $E_C$, which depends on the capacitance values, as well as the voltage distribution between the two junctions, which is determined by the capacitance ratio, $V_1/V_2 = C_2/C_1$ (see Fig. 1).[1,7] The ratio between the tunneling resistances may affect the degree of QD charging during the tunneling process through the DBTJ.[8]

Previously we have performed tunneling spectroscopy measurements on InAs nanocrystals linked to gold by hexane-dithiol molecules realizing a capacitively highly asymmetric DBTJ ($C_2/C_1 \sim 10$), where CB (VB) states appeared at positive (negative) bias.[1] Single electron charging peaks were observed, from which two and six-fold degenerate *s* and *p*-like levels were identified, and directly imaged in a subsequent work on InAs/ZnSe core/shell QDs.[11] The observation of QD charging indicated that the tunneling rate $\Gamma_2 \propto 1/R_2$ was on the order of $\Gamma_1 \propto 1/R_1$. Otherwise, for positive sample bias, an electron tunneling from the tip to the QD would escape to the substrate before the next electron could tunnel into the QD (and an equivalent process would occur for negative bias). Consequently, merely resonant tunneling through the QD states without charging would take place. By varying the tip-QD distance we were able to modify the voltage division between junctions up to the distance that allowed us to obtain meaningful (well above the noise level) tunneling spectra. Similarly, Bakkers and Vanmaekelbergh also reported an STM study of CdS and CdSe QDs, focusing on the role of voltage division.[3] However, we could not observe a clear transition to a state of charging-free resonant tunneling. This may be due to the large $R_2$ value resulting from the linker molecules that anchored the QDs to the substrate.[1] In this paper we show that upon reducing $R_2$, by working without linker molecules, charging-free resonant tunneling, as well as a transition back to tunneling accompanied by QD charging, can be achieved for a single QD by controlling $\Gamma_2$.

InAs nanocrystals capped by organic ligands were prepared using colloid chemistry technique[12] and deposited on freshly cleaved highly oriented pyrolitic graphite (HOPG). Freshly cleaved HOPG surfaces are atomically flat as compared to the Au films we previously used as substrates in our experiments, facilitating the STM observation of QDs. However, the unlinked QDs were very mobile on the surface even at 4.2 K and were partially immobilized only near steps on the HOPG surface,[13] as depicted by Fig. 1. This made both the imaging and spectroscopic measurements relatively difficult. Distorted images were frequently observed due to QD dragging by the tip, whereas in the spectroscopic measurements, the QD could diffuse away before establishing optimal $I_S$ and $V_S$ values.

In Fig. 2a we plot a tunneling spectrum measured on an InAs QD, 2 nm in radius, using the tunneling configuration portrayed in the right inset of Fig. 1 (solid curve). For comparison, we also plot a representative spectrum measured on a QD of similar radius, but anchored to a gold substrate via linker molecules (dashed line). The tunneling spectra were obtained either directly using lock-in technique or by numerical differentiation of measured I-V curves, yielding similar results. Both spectra exhibit a gap in the density of states around zero bias, associated with the QD energy band-gap, and peaks at positive (negative) bias reflecting the CB (VB) states. However, there is a profound difference between these two spectra. In the spectrum measured in the QD/linker-molecule/Au geometry, resonant tunneling accompanied by QD charging is clearly seen, exhibited (in the CB) by the doublet of peaks that is followed by a higher order multiplet. As discussed previously,[1] the doublet corresponds to tunneling through the two-fold degenerate *s*-like CB state (denoted $1S_e$) with the spacing assigned to the single electron charging energy, while the higher order multiplet is assigned to tunneling through the *p*-like state ($1P_e$). The charging multiplets are absent in the spectrum measured in the QD/HOPG geometry, and each multiplet is replaced by a single (somewhat broadened) peak, indicating charging-free resonant tunneling through the *s* and *p*-like CB states. We note that the charging multiplets were absent even when the peaks did not exhibit significant broadening (e.g., the *s* peak in the upper curve of Fig. 3). A similar behavior is seen also for the more complex VB. We attribute the difference between the two spectra to the different tunneling rate ratios $\Gamma_2/\Gamma_1$, achieved in either of the DBTJ's. A significantly lower tunnel barrier of the QD-substrate junction is expected in the QD/HOPG configuration.

To confirm this interpretation we performed theoretical simulations using the "Orthodox Model" for single electron tunneling[7] modified to account for the effect of a discrete QD level spectrum.[9] The solid and dashed theoretical curves presented in

*Katz et al.* 4

Fig. 2b were calculated assuming the same two-fold and six-fold degenerate (*s* and *p*) CB levels, and two four-fold degenerate VB states.[1,14,15] The capacitance values were also kept the same for the two curves, $C_1 = 0.1$ aF and $C_2 = 1.1$ aF, resulting in a ~90% voltage drop on the tip-QD junction and $E_C \sim 100$ meV. The two curves differ in the ratio between the tunneling rates: $\Gamma_2/\Gamma_1 = 1$ and 10 for the dashed and solid curves, respectively (only the ratio is relevant for our discussion). The dashed curve shows strong charging multiplets, typical for resonant tunneling taking place along with QD charging. The solid curve, on the other hand, exhibits only a signature of charging effect (e.g., one small charging peak in the *p* multiplet), which vanished for $\Gamma_2/\Gamma_1$ larger than 100. It is evident that the curve for $\Gamma_2/\Gamma_1 = 1$ resembles the experimental spectrum obtained for the QD/linker-molecule/Au system, while the $\Gamma_2/\Gamma_1 = 10$ curve better corresponds to the QD/HOPG configuration, consistent with our interpretation above. Note also that the apparent *s-p* level separation (both in theory and experiment) is smaller for the QD/HOPG configuration, due to the absence of charging contribution.

Furthermore, a transition from charging-free tunneling to resonant tunneling in the presence of charging is demonstrated by Fig. 3, for the QD-HOPG system. Here we plot two tunneling spectra acquired on the same QD of radius 2.5 nm, with different tip-QD separations. The dashed curve was measured with $V_S = 1.5$ V and $I_S = 0.1$ nA, while the solid curve was taken with $I_s = 0.8$ nA, moving the tip closer to the QD. There are two marked differences between these two spectra. First, the apparent gap in the density of states around zero bias is larger for the curve measured with the tip closer to the QD. This is attributed to the effect of voltage division between the two junctions. In our measurements $C_1$ is smaller than $C_2$, therefore the applied voltage $V_B$ largely drops on the tip-QD junction and tunneling through the discrete QD levels is onset in this junction. Hence, the apparent level spacing in the tunneling spectra is larger than the real level spacing by a factor of $V_B/V_1 = (1+C_1/C_2)$. Therefore, upon reducing the tip-QD distance, $C_1$ increases and so does the measured gap. The second difference is even more profound. In the dashed curve, a doublet is observed at the onset of tunneling into the CB, in contrast to a corresponding single peak seen in the solid curve. The second peak in the dashed curve cannot be associated with the *p* state since the apparent *s-p* separation must be larger here as compared to the solid curve due to the effect of voltage division, while the observed spacing is smaller. The peak spacing within this doublet is 170 meV, comparable to $E_C$ values measured for InAs QDs of similar size.[1] Hence, we attribute this doublet to single electron charging of the $1S_e$ level. As the tip approaches the QD, $\Gamma_1$ increases towards the value of $\Gamma_2$ and thus the process of resonant tunneling becomes accompanied by QD charging. The tunneling current exceeded the saturation value of detection at a voltage that did not allow us to check whether the *p*-level also becomes charged.

To address in greater detail the effect of voltage division within the DBTJ, the VB-CB and *s-p* gaps were extracted from the solid curve in Fig. 3, yielding 1.56 and 0.52 eV respectively. Both values are 1.25 times larger than those obtained previously[1] for InAs QDs of similar size. Therefore, even this curve was still measured with non-negligible voltage drop across the QD-HOPG junction. The spectra presented in solid lines in Fig. 4 were obtained on three different QDs, with the tip further retracted from the QD (using a typical setting of $V_S = 2$ V and $I_S = 50$ pA). Here, the measured VB-CB gaps (1.57, 1.37, and 1.2 eV) and *s-p* level separations (0.49, 0.43, and 0.32 eV) for the nanocrystals of radii 1.8, 2.5, and 3.4 nm, respectively, are in relatively good agreement with Ref. 1, and exhibit the expected quantum-size effect. The dashed curve at the bottom of the figure was acquired on the 3.4 nm QD, but with $I_S = 0.1$ nA. Again, the *measured* gaps broaden due to the increase of $C_1/C_2$, this time without the onset of QD charging, i.e., $\Gamma_1$ was still too small as compared to $\Gamma_2$. The setting of $I_S$ to lower values allowed us to observe (for the larger QDs) a third peak at positive bias before current saturation. This peak may be related to the next CB state, presumably $1D_e$,[16] which could not be detected in our previous work since, due to the effect of charging, it was pushed out to voltages beyond the limit of current saturation or the onset of field emission.

In summary, we have demonstrated the effect of varying the DBTJ parameters on the measured tunneling spectra. The capacitance ratio between the junctions was found to primarily affect the apparent level spacing, whereas by varying the ratio between the tunneling rates, a control over the degree of QD charging during the tunneling process is achieved.

We thank Y-M. Niquet for helpful discussions. This work was supported by the Israel Science Foundation founded by the Israel Academy of Science, the BIKURA foundation, and by Intel-Israel.

Figures and Captions:

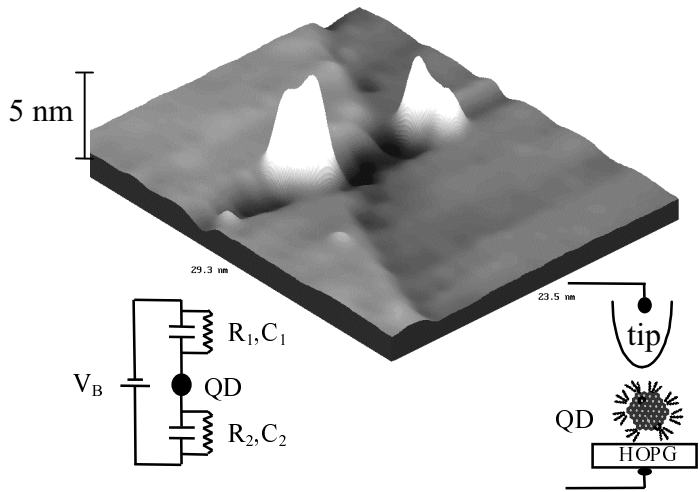

**Figure 1:** A 30nm x 30nm topographic STM showing two single InAs QDs positioned near a monolayer step on HOPG. By positioning the STM tip above the QD, a DBTJ configuration is realized (shown schematically in the right inset), as depicted by the equivalent circuit in the left inset.

... 

*Katz et al.* 8

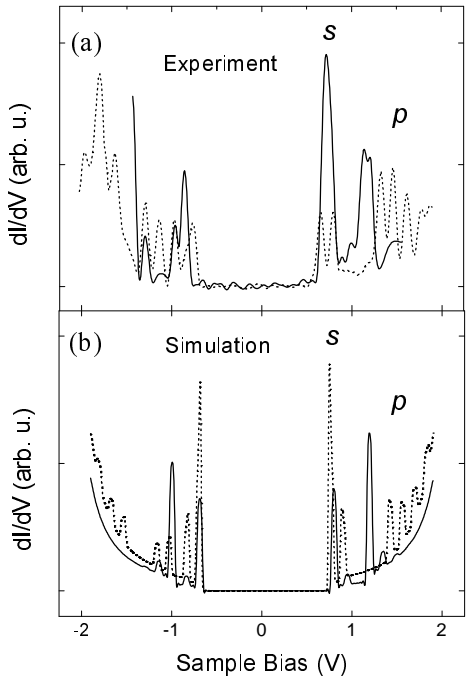

**Figure 2:** (a) Tunneling spectra measured on InAs QDs ~ 2 nm in radius. The solid curve was measured in the QD/HOPG geometry, and the dashed curve in the QD/linker-molecule/Au geometry. (b) Calculated spectra showing the effect of tunneling-rate ratio. The dashed and solid curves were calculated with $G_2/G_1 = 1$ and 10, respectively, see text.

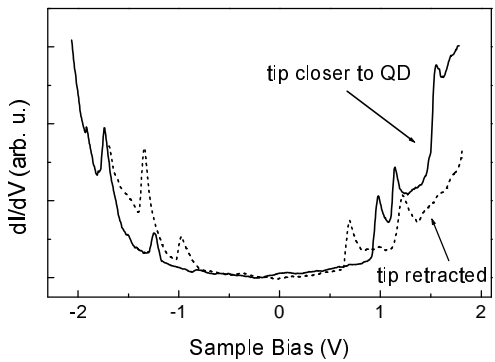

**Figure 3:** Tunneling spectra measured on a single QD ($r \sim 2.5$ nm) with two different tip-QD separations, exhibiting effects on both the apparent QD level spacing and charging, see text.

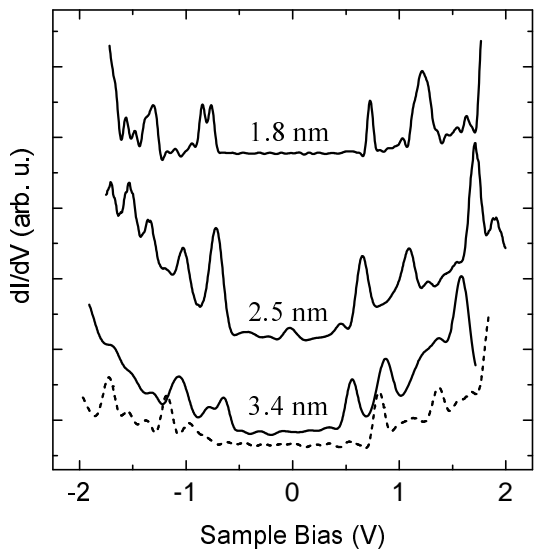

**Figure 4:** Size evolution of tunneling spectra of InAs QDs on HOPG. The QD radii are denoted in the figure. For clarity, the spectra are shifted along the bias axis to center the measured gaps at zero bias.